\documentclass[fleqn,usenatbib]{mnras}

\usepackage[T1]{fontenc}

\DeclareRobustCommand{\VAN}[3]{#2}
\let\VANthebibliography\thebibliography
\def\thebibliography{\DeclareRobustCommand{\VAN}[3]{##3}\VANthebibliography}

\usepackage{graphicx}	% Including figure files
\usepackage{amsmath}	% Advanced maths commands
\usepackage{amssymb}	% Extra maths symbols
\usepackage{color}
\usepackage{changes}
\title[A Pilot Study of Interplanetary Scintillation with FAST]{A Pilot Study of Interplanetary Scintillation with FAST}

\author[Li-Jia Liu et al.]{
Li-Jia Liu,$^{1,2}$\thanks{E-mail: \href{mailto:ljliu@nao.cas.cn}{ljliu@nao.cas.cn}}
Bo Peng,$^{1}$\thanks{E-mail: \href{mailto:pb@nao.cas.cn}{pb@nao.cas.cn}}
Lei Yu,$^{1,2}$\thanks{E-mail: \href{mailto:yulei@nao.cas.cn}{yulei@nao.cas.cn}}
Ye-Zhao Yu,$^{3}$
Ji-Guang Lu,$^{1}$
Bin Liu,$^{1}$\thanks{E-mail: \href{mailto:bliu@nao.cas.cn}{bliu@nao.cas.cn}}
O. Chang,$^{4}$
M. M. Bisi,$^{4}$
\and the FAST Collaboration$^{1}$
\\
$^{1}$CAS Key Laboratory of FAST, National Astronomical Observatories, Chinese Academy of Sciences, Beijing 100101, China\\
$^{2}$University of Chinese Academy of Sciences, Beijing, 100049, China\\
$^{3}$School of Physics and Electronics, Qiannan Normal University for Nationalities, Duyun, 558000, China\\
$^{4}$RAL Space, United Kingdom Research and Innovation - Science and Technology Facilities Council - Rutherford Appleton Laboratory, \\
Harwell Campus, Oxfordshire, OX11 0QX, UK
}

\date{Accepted 2021 April 20. Received 2021 April 19; in original form 2020 October 01}

\pubyear{2015}

\begin{document}
\label{firstpage}
\pagerange{\pageref{firstpage}--\pageref{lastpage}}
\maketitle

\begin{abstract}

Observations of Interplanetary Scintillation (IPS) are an efficient remote-sensing method to study the solar wind and inner heliosphere. From 2016 to 2018, some distinctive observations of IPS sources like 3C\,286 and 3C\,279 were accomplished with the Five-hundred-meter Aperture Spherical radio Telescope (FAST), the largest single-dish telescope in the world. Due to the 270-1620\,MHz wide frequency coverage of the Ultra-Wideband (UWB) receiver, one can use both single-frequency and dual-frequency analyses to determine the projected velocity of the solar wind. Moreover, based on the extraordinary sensitivity owing to the large collecting surface area of FAST, we can observe weak IPS signals. With the advantages of both the wider frequency coverage and high sensitivity, also with our radio frequency interference (RFI) mitigation strategy and an optimized model-fitting method developed, in this paper, we analyze the fitting confidence intervals of the solar wind velocity, and present some preliminary results achieved using FAST, which points to the current FAST system being highly capable of carrying out observations of IPS.
\end{abstract}


%%%%%%%%%%%%%%%%%%%%%%%%%%%%%%%%%%%%%%%%%%%%%%%%%%

%%%%%%%%%%%%%%%%% BODY OF PAPER %%%%%%%%%%%%%%%%%%
\section{Introduction}

While coming through the solar wind plasma, the radio signal from a distant compact radio source is scattered by the density inhomogeneities of the solar wind, consequently a random diffraction pattern is observed at Earth, this phenomenon is known as Interplanetary Scintillation (IPS) \citep{hewish1964interplanetary}. In turn, these ground-based IPS observations can be used to infer the physical properties of the solar wind. Nowadays, as the space-exploration and technologies are growing, the solar-activity monitoring and the space-weather forecasting are becoming more critical. Although the solar wind can be observed directly using an {\it in-situ} satellite/spaceceaft with relatively-high accuracy, the ground-based observations of IPS provide a more economic method to obtain the information of the solar wind, such as velocity as well as the structure of a sub-arcsecond scale compact radio source \citep{hewish1969radio,1972JGR....77.4602A}.
Another advantage of observations of IPS is that the solar wind can be monitored with a longer time scale and a more flexible spatial range out of the ecliptic plane. Besides solar physics, observations of IPS can also be used to study space weather.\\

Since the IPS phenomenon was discovered in the 1960s \citep{1964Clarke, hewish1964interplanetary}, many countries including the UK, USA, Japan, India and, Russia, have undertaken IPS studies. Some astronomical instruments were built exclusively for IPS observations, like the multi-station system in Japan \citep{asai1995multi}. Another successful example is called the Ooty Radio Telescope (ORT) located in India \citep{manoharan1990determination}. The two facilities are both based on a parabolic cylinder design using a central observing frequency of 327\,MHz. Since observations of IPS are an effective way to study heliospehric physics, some new facilities for IPS studies have been built or are going to be built, like the Mexican Array Radio Telescope (MEXART) \citep{2010SoPh..265..309M, 2016AdSpR..57.1307C} in Mexico, and a new IPS telescope system to be constructed in Inner Mongolia, China, which is an array with three 140*40\,m cylinder antennas at Mingantu, together with two 16\,m parabolic antennas at Abaga and Keshiketeng \citep{2018SunGe..13..153Y}. Furthermore, other radio telescopes, like the Murchison Widefield Array (MWA) \citep{2015ApJ...809L..12K}, Low Frequency Array (LOFAR) \citep{2013SoPh..285..127F}, and Square Kilometre Array (SKA) \citep{2015aska.confE.169N,2019AdSpR..63.1404N} have also adopted IPS investigations as one of their scientific goals.\\

Observations of IPS can be conducted either by a single station or by multi stations \citep{1972JGR....77.4602A, coles1978solar, 2007ChJAA...7..712Z, 2010SoPh..261..149B, 2010SoPh..265...49B}. The four-station system in Japan is a representative of the multi-station system, and it can measure the radial solar wind projection \citep{asai1995multi}. Due to some practical reasons, most IPS facilities have adopted the single station system, like the ORT \citep{2010SoPh..265..137M}, MWA \citep{2018MNRAS.473.2965M, 2018MNRAS.474.4937C, 2018MNRAS.479.2318C} and the 25m radio telescope in Urumqi, Xinjiang, China \citep{liu2010observations}. For the single station observations, two analysis modes are available, dependent on the system observing capabilities, namely the single-frequency (SSSF) analysis mode, and the dual-frequency (SSDF) analysis mode \citep{2007ChJAA...7..712Z}. In the single-frequency analysis, the solar wind velocity can be calculated from a multi-parameter model fitting or from the characteristic frequencies of the power spectrum. As for the dual-frequency analysis, the solar wind velocity can be drawn from the first zero-crossing frequency of the normalised co-spectrum (N.C.S).\\

In China, IPS observations were started in 1990s with the Miyun Synthesis Radio Telescope (MSRT) in Miyun station, which was run by Beijing Astronomical Observatory (BAO), Chinese Academy of Sciences and now is called National Astronomical Observatories, Chinese Academy of Sciences (NAOC) \citep{1993Ma, 2001IAUS..203..580Z, 2001Ap&SS.278..189W}, which is an array of 28 single dishes. The effective area of MSRT is equal to a 47\,m single dish, with the observing frequency of 232\,MHz.\\

An IPS quasi-regular observation system was established with the 25\,m radio telescope in Urumqi \citep{liu2010observations} in 2008, with a central observing frequency of 1.4GHz. It is the first IPS observation system launched in a single dish in China. The observing frequency of the telescopes in Ooty and Solar-Terrestrial Environment Laboratory, Nagoya University (STELab, which is now called ISEE) are both 327MHz. Since IPS observations can capture distance information corresponding to frequencies, while observing IPS sources at higher frequencies, one can obtain the information relatively closer to the Sun \citep{1983A&A...123..191S}, therefore the Urumqi telescope can obtain information on approaching position less than 18 solar radii\citep{2007ChJAA...7..712Z}.\\

In 2006 a mega-science project in China was launched, called the Meridian Space Weather Monitoring Project (Meridian Project for short). There is a sub-system of this project to develop ground-based IPS observations, which is configured on the 50\,m telescope in Miyun station run by NAOC. There are two dual-frequency receivers available in this system, one is centered at 327/611\,MHz and the other is centered at 2300/8400\,MHz. The bandwidths of the four central frequencies are 40, 40, 300 and 800\,MHz respectively. This IPS facility aimed to obtain the solar wind velocity and scintillation index information for space weather forecasting \citep{2012RAA....12..857Z}. In the near future, an IPS facility will be built in Inner Mongolia for the National Meridian Project 2  \citep{2018SunGe..13..153Y}, which is a three-station system with observing frequencies centered at 327\,MHz, 654\,MHz, and 1.4\,GHz.\\

The Five-hundred-meter Aperture Spherical radio Telescope (FAST) was constructed successfully in Pingtang country, Guizhou province of China in 2016. In the commissioning phase, some preliminary IPS observations have been carried out. The observation parameters of FAST and ORT  {\citep{2000Oberoi} are given in Table~\ref{tab:1}}. According to Table~\ref{tab:1}, to observe a target source of 1\,Jy with an integration time of 1\,s with a typical bandwidth of 4\,MHz, FAST can achieve a signal-to-noise ratio (S/N) of 95, while the ORT reaches a S/N of 25 with the same settings. Based on the high sensitivity of FAST and the wide frequency coverage of the Ultra-Wideband (UWB) receiver, we can observe weak IPS phenomena within a short time, and analyze the confidence intervals associated to solar wind velocity, which is reported in section 3. \\

The IPS observation with FAST is introduced in section 2. The data reduction is presented in section 3. Discussions and concluding remarks are then presented in section 4.

\section{IPS Observation with FAST}

\begin{figure}
	\includegraphics[width=\columnwidth]{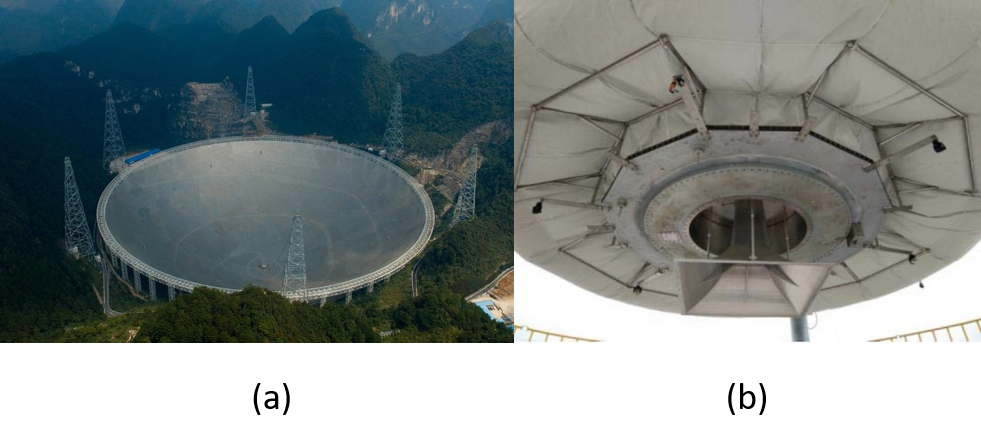}
    \caption{The picture of FAST (a), equipped with the Ultra-Wideband (UWB) receiver (b)}
    \label{fig:1}
\end{figure}

Since 1993, astronomers and engineers from different countries including China have been involved in an ambitious international science project, later referred  as the Square Kilometre Array (SKA) with a collecting area of one square kilometer, while the location of FAST was a potential candidate site of the SKA \citep{1997hsra.book..278P, 2000SPIE.4015...45P} as the Chinese SKA program \citep{2002RaScB.300...12P}. FAST is now the largest single-dish radio telescope in the world Fig.~\ref{fig:1} (a), with a unique sensitivity to carry out some revolutionary scientific goals, like surveying neutral hydrogen in the universe, detecting the faint and rare types of pulsars, looking for the first shining stars, etc. \citep{2000pras.conf...25P, 2011IJMPD..20..989N}.\\

After the first light of FAST observations in September 2016, FAST entered into its commissioning phase. Some calibrators used in the FAST testing phases are IPS sources such as 3C\,286 and 3C\,279. As a benefit of the big collecting area, FAST has an extraordinary sensitivity, which is shown in Table~\ref{tab:1}. Compared to other telescopes, the unique high sensitivity of FAST improves the capability of weak signal detection in IPS observations. As a result, a series of IPS experiments were completed successfully by this giant radio telescope.\\

\begin{table}
	\centering
	\caption{Key observation parameters of ORT and FAST. $T_{sys}$ for FAST is provided for 327\,MHz for direct comparison with ORT.
	}
	\label{tab:1}
	\begin{tabular}{lccr} % four columns, alignment for each
		\hline
		Facility & $T_{sys}$ &  $A_{eff}$ \\

                 & (K)&($m^{2}$)\\
		\hline
		ORT & 150  & 8000\\
		FAST & 70 &  49500\\
		
		\hline
	\end{tabular}
\end{table}

The aperture diameter of FAST is 500\,m with the frequency coverage from 70\,MHz to 3\,GHz. The high-frequency end will be extended to 8\,GHz in future upgrades \citep{2011IJMPD..20..989N}. From July 2016 to May 2018, before the installation of the 19-beam receiver, the Ultra-Wideband (UWB) receiver Fig.~\ref{fig:1} (b) was used, by which all IPS data were observed. This receiver covers a frequency range of 270-1620\,MHz. \\

Like other observing systems, the IPS observing system in FAST consists the dish, UWB receiver, digital back-end, and the data processing pipeline. The UWB receiver mounted on FAST has a large bandwidth ratio of 1:6. The wide coverage feature can effectively reduce the number of receivers. The entire bandpass of the Reconfigurable Open Architecture Computing Hardware (ROACH) for data collection is divided into multi-bands, the data quality is flexibly controlled and can be explicitly displayed and checked. In order to obtain both on-source and off-source data, the tracking mode of FAST was taken for our IPS observations.\\

The FAST UWB receiver has a digital acquisition back-end that includes two polarizations designed for pulsar observation, which is suitable for IPS observations. The sampling interval of 0.1\,ms is sufficient for IPS studies. The data acquisition of the UWB receiver is separated equally into two bands. The low-frequency band covers 270-800\,MHz and the high-frequency band covers 1200-1620\,MHz. Both bands are then divided into multi-channel, and each channel covers a frequency window of 0.25\,MHz.

\section{Data Reduction}
From late 2016 to May 2018, a series of experimental IPS observations were carried out with FAST. After some system testing, debugging and calibration, the IPS observation was firstly making use of the calibration data from FAST to monitor the solar wind. The sources observed, typical sampling rate, the dates and the duration time of the observations are given in Table~\ref{tab:2}\\

\begin{table}
	\centering
	\caption{Details of observations.
	}
	\label{tab:2}
	\begin{tabular}{lccr} % four columns, alignment for each
		\hline
		Source  & Dates  & Sampling rate & Duration time \\
                & (2017)& (\,ms) & (\,min)
                 \\
		\hline
		3C\,286  & 11/13, 11/17 & 0.1 & 10\\
		3C\,279  & 11/13, 11/18 & 0.1 & 10\\
		
		\hline
	\end{tabular}
\end{table}

Equipped with the wideband receiver, observations of IPS with FAST can be conducted in both SSSF and SSDF analyses methods simultaneously, which helps us to get the solar wind velocity ($V$), the anisotropic axial ratio (AR), which is the ratio of the major to the minor axis of the density irregularity {\citep{1998JGR...103.6571Y}}, the spectral index of electron density fluctuation ($\alpha$) and source size ($\theta_0$) with high precision. During the commissioning phase, the high-frequency band system was not very stable.  therefore our discussions are focused on the low-frequency band data.\\

In this section, the data reduction pipeline and data processing procedure is introduced.  Including the choice of IPS observing frequency, bandwidth, the length of time series, and a model-fitting method to obtain the solar wind velocity.

\subsection{Data Reduction Pipeline}
\label{sec:Data Reduction Pipeline} % used for referring to this section from elsewhere

For IPS data processing, the radio frequency interference (RFI) in the data recorded by the UWB receiver are first removed, to form a new time series which is integrated both on time and frequency.\\

\begin{figure}
	\includegraphics[width=\columnwidth]{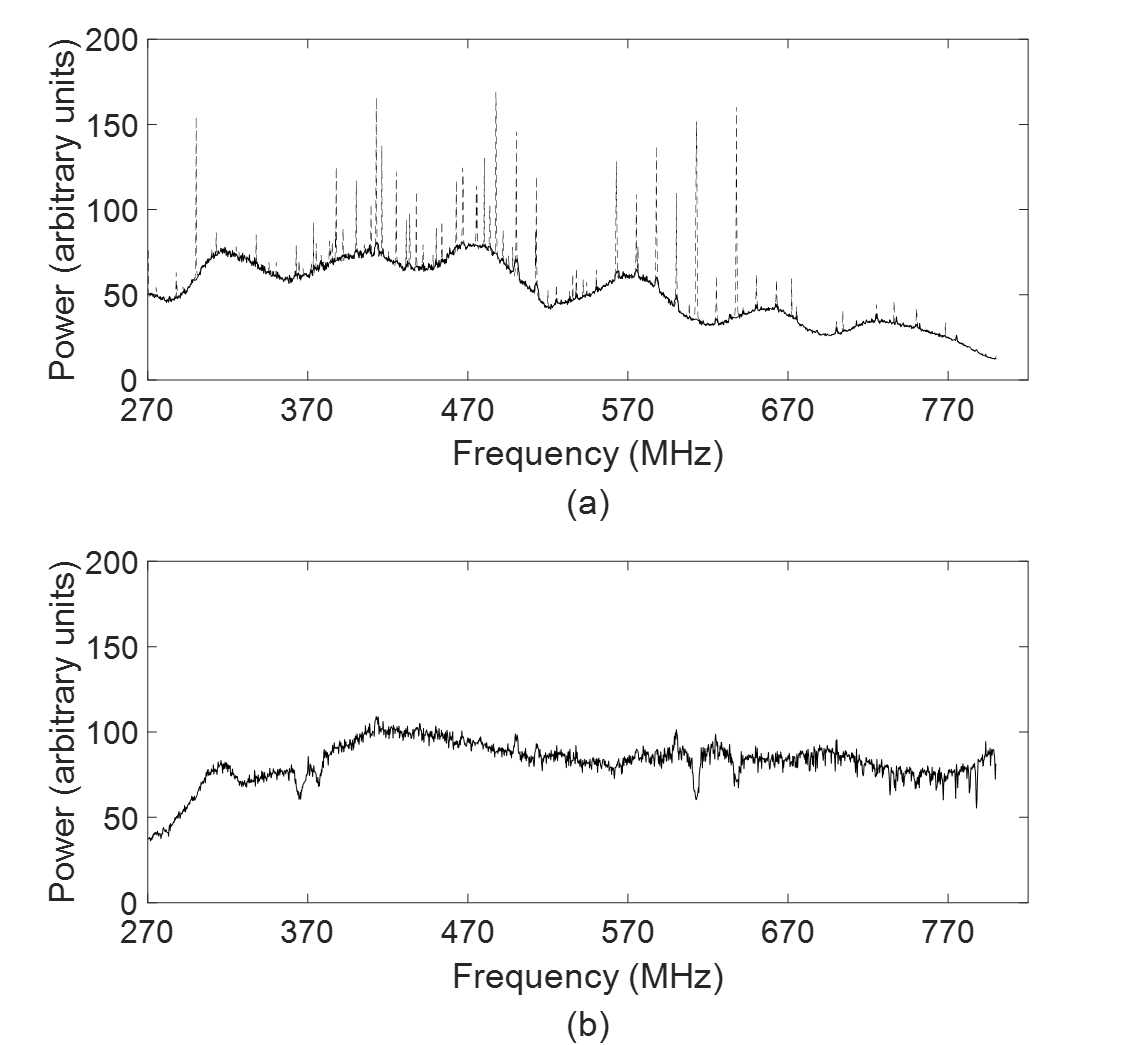}
    \caption{The frequency spectrum of the FAST Ultra-Wideband (UWB) receiver showing a standing-wave pattern. The frequency coverage of this spectrum is 270-800\,MHz, with a duration of 20\,ms. The target source was 3C\,286, and was observed on 13 November 2017. (a) The dashed and solid lines are the frequency-spectrum before and after RFI removal respectively. (b)The normalized spectrum which has eliminated the systematic instrumental response between different frequency.}
    \label{fig:2}
\end{figure}

Fig.~\ref{fig:2} demonstrates a data sample of the 270-800\,MHz frequency-spectrum collected by the UWB receiver with a duration time of 20\,ms. The target source was 3C\,286, and was observed on 13 November 2017. In Fig.~\ref{fig:2} (a) the dashed and solid lines are the frequency-spectrum before and after RFI removal respectively. Fig.~\ref{fig:2} (b) is the normalized spectrum which has eliminated the systematic instrumental response between different frequency. The broad-band fluctuation in the bandpass is probably due to standing waves, while the strong narrow-band lines are caused by the RFI. It is obvious that, after removing the RFI, the interferences through the whole bandpass are significantly suppressed. For traditional receivers, the acquisition data are summed throughout the whole bandwidth, which will cause some uncertainty to the collected data, therefore, it is essential to remove the RFI. However, the multi-channel design in the FAST UWB receiver in a radio-quiet location allows us to identify the channels with RFI and removed it more effectively.\\

A general expression of the telescope observed signal is:
\begin{equation}
    S_{obs}=S_{sig}+\varepsilon+S_{RFI}.
	\label{eq:1}
\end{equation}

Where $S_{obs}$ is observed signal, $S_{sig}$ is the signal-of-interest(SOI), $\varepsilon$ is the additive background white Gaussian noise and $S_{RFI}$ is RFI. We assume that the RMS of the SOI with noise is different from the RFI. As a result of the baseline of the frequency-spectrum is very complicated, we use the rate of change of the SOI rather than the spectrum itself to identify SOI and RFI. Then Random sample consensus (RANSAC) is applied to remove RFI. The main steps are: 1) Initializing the hyperparameters (inner point ratio and tolerance boundary) and fitting model selection (a line model). 2) Convolving the spectrum with Laplacian of Gaussian filter. 3) Fitting the line model and identify the inner and outer data. 4) Doing average interpolation for the outer data. 5) Repeating the steps 2 to 4 until the RMS of spectrum convergence. The narrow-band RFI can be correctly removed by this method, and based on that, eliminated the relatively wide-band RFI can be achieved by iteratively the method.\\

\begin{figure}
	\includegraphics[width=\columnwidth]{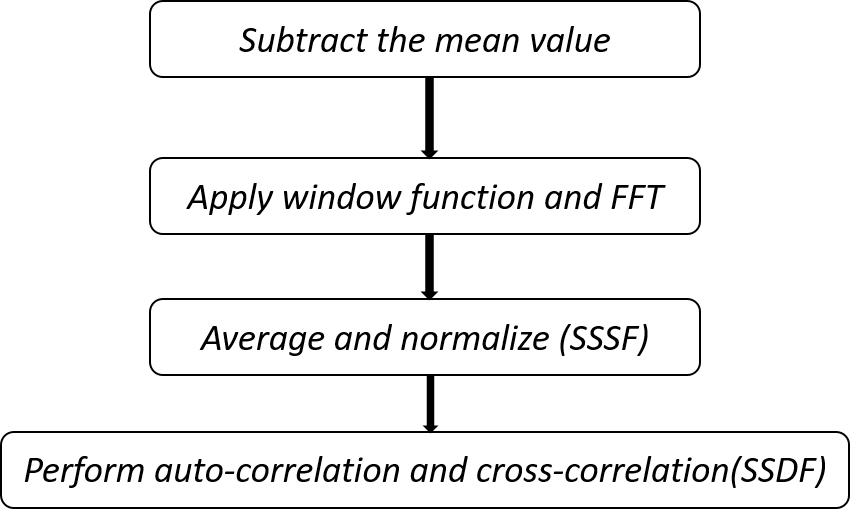}
    \caption{The data reduction pipeline for SSSF and SSDF analyses.}
    \label{fig:3}
\end{figure}

The bandwidth of each point of these new data is 10\,MHz with an integration time of 20\,ms. The data reduction pipeline Fig.~\ref{fig:3} is described as follows: 1) The new data set is then divided into some sub-set, with 512 data points and a duration time of about 10\,s. Each sub-set is subtracted by its average since the IPS phenomenon only relates to the flux variance. 2) Apply the fast Fourier transform (FFT) to obtain the power spectrum, from which one can extract variance information in frequency scale. The data sample is smoothed using Hanning Window as the Window function so as to reduce the cut-off effects of the FFT. 3) Average the spectrum of 2 sub-set and normalize it to form the SSSF power spectrum of a time duration of 20\,s. There is some observational evidence for AR to be close to unity when the solar elongation is larger than $15^\circ$ \citep{coles1978solar, 1996AIPC..382..366Y}, that means the low-frequency part of the SSSF power spectrum is almost flat. The elongation of 3C\,286 on 13 November 2017 was beyond $30^\circ$. So in the following of the paper, the data points of 3C\,286 below 0.7\,Hz are set to be the same value. The data processing steps 1) to 3) are applied subsequently to the SSSF analysis. The last step is the SSDF analysis based on {\citet{1983A&A...123..191S}}, which takes the FFT outputs from step 2) to perform auto-correlation and cross-correlation for each dual-frequency pair.\\

\subsection{Data Processing Results}

The IPS observation was intended to monitor the solar wind and to calculate the velocity and scintillation index. The UWB receiver is capable of taking data over very wide bandwidths which can be split up and analysed very flexibly thus making it possible to perform both SSSF and SSDF analyses on the same data set simultaneously. Furthermore, any observing frequency bands could be chosen flexibly. Since there was no extra observation time to be applied, calibrators like 3C\,286 and 3C\,279 are chosen for our IPS study.\\

The on- and off-source raw data of 3C\,286 and 3C\,279 observed on 13 November 2017, are shown in Fig.~\ref{fig:5} (a) and (b) respectively. The observing frequency is centered at 305\,MHz. The fluctuation levels for on- and off-source observations have a significant difference as shown in Fig.~\ref{fig:4} (a) and (b). There is obvious discrimination as shown in the two rectangles in Fig.~\ref{fig:4} (a). While in Fig.~\ref{fig:4} (b), the fluctuations in the two rectangles are at a similar level. {Which can also be reflected by the RMS values. So the scintillation level of 3C\,286 that day was stronger than that of 3C\,279. In the following, the results derived from the observation on 3C\,286 are discussed as an example.

\begin{figure}
	\includegraphics[width=\columnwidth]{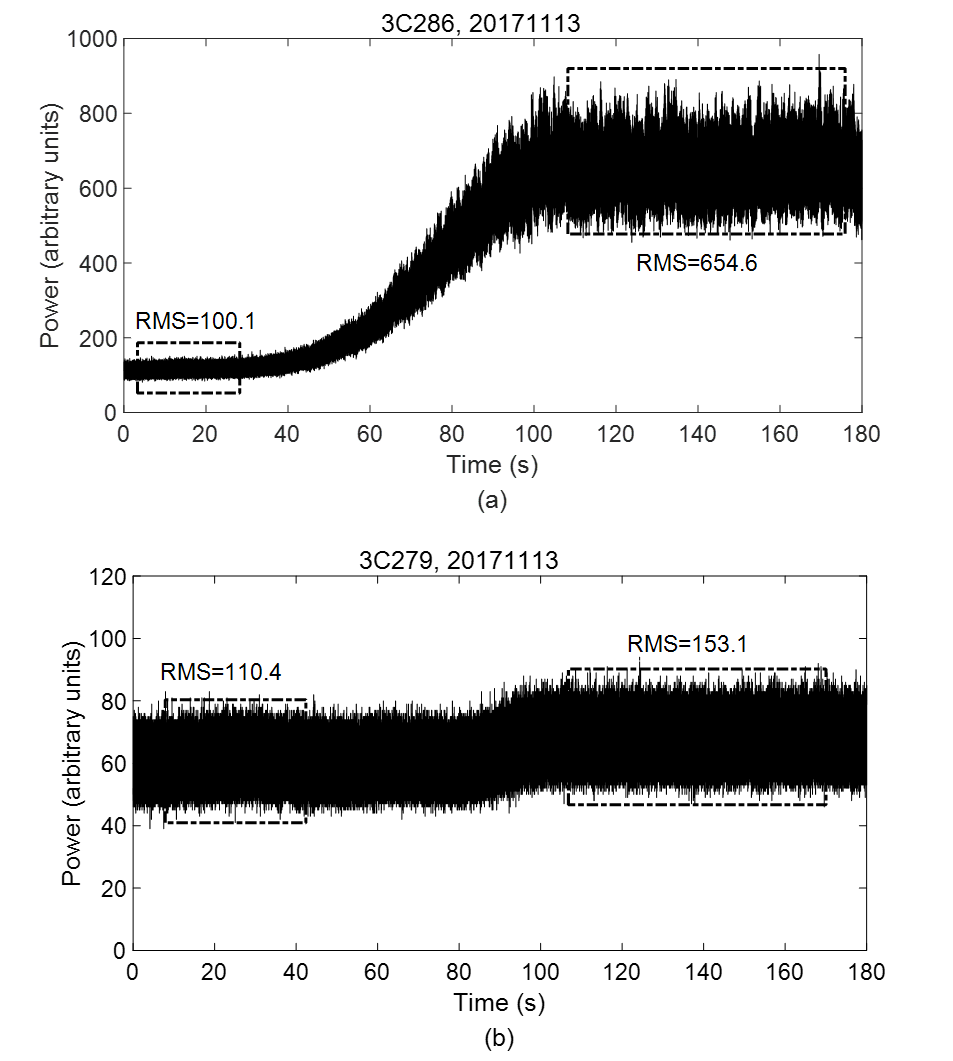}
    \caption{The off-(on) raw data of 3C\,286 (a) and 3C\,279 (b) observed at 305 MHz on 13 November 2017, respectively, the four rectangles present the fluctuations of on-source and off-source. }
    \label{fig:4}
\end{figure}

\begin{figure}
	\includegraphics[width=\columnwidth]{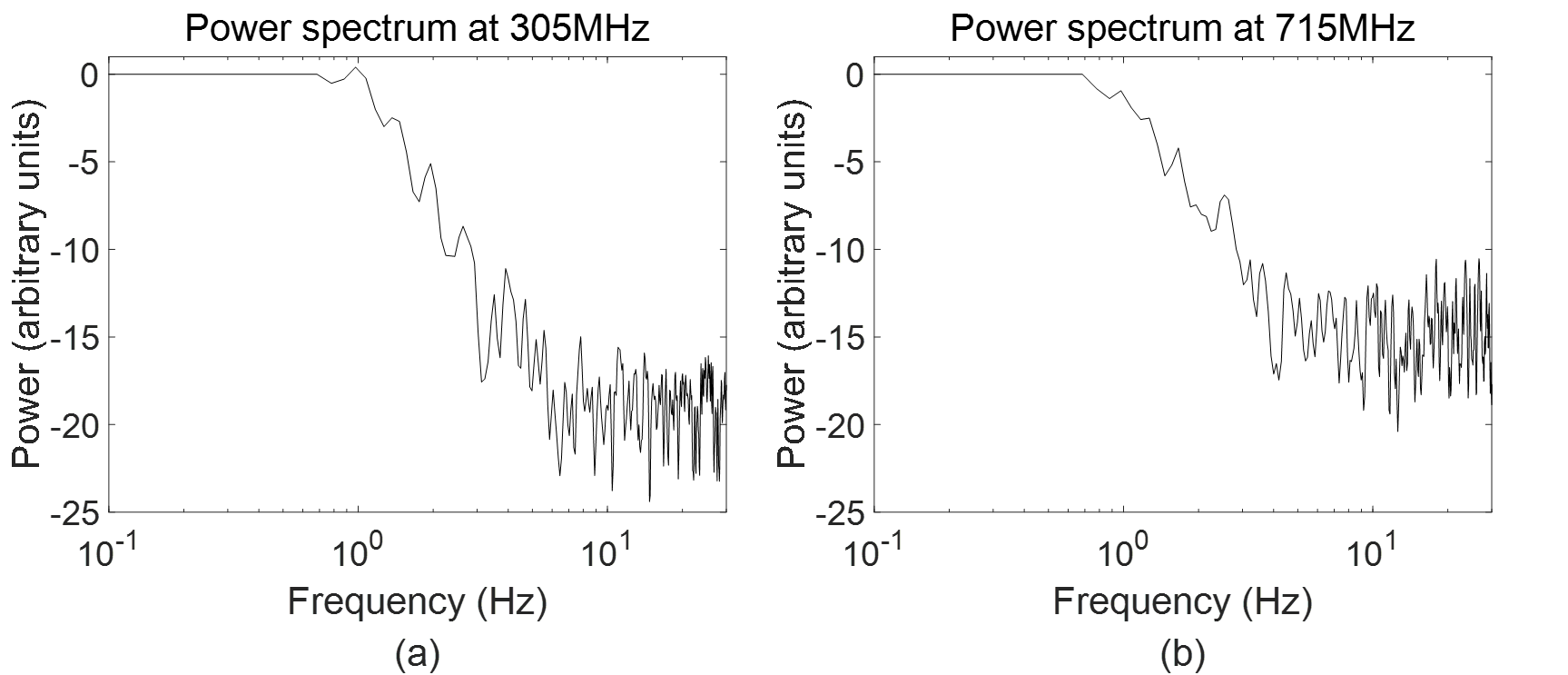}
    \caption{SSSF analysis mode power spectra for 3C\,286 at 305\,MHz and 715\,MHz on 13 November 2017. }
    \label{fig:5}
\end{figure}

Fig.~\ref{fig:5} demonstrates the SSSF analysis mode power spectrum of the source 3C\,286 observed on 13 November 2017. The central frequencies for Fig.~\ref{fig:5} (a) and (b) are 305\,MHz and 715\,MHz, respectively, and each of which has a bandwidth of 10\,MHz. Fig.~\ref{fig:5} clearly shows that the power spectrum at 305\,MHz has a higher scintillation level than the spectrum at 715\,MHz. According to equation~(\ref{eq:2}), the scintillation index $m$ \citep{cohen1967interplanetary} is 0.09 for the low frequency and 0.08 for the high frequency, where $C_{on}$ ($C_{off}$) is the average intensity of the on-source (off-source) signal, and $\sigma^2_{on}$ ($\sigma^2_{off}$) is the square of the RMS of intensity scintillation. \\

\begin{equation}
    m=\frac{\sqrt{\sigma^2_{on}-\sigma^2_{off}}}{C_{on}-C_{off}}.
	\label{eq:2}
\end{equation}

\subsection{Model-fitting}

The solar wind velocity from SSSF analysis can be obtained either by 1) the characteristic frequency called Fresnel knee frequency or 2) by model fitting via some parameters. In equation~(\ref{eq:3})\citep{1983A&A...123..191S}, V stands for the solar wind velocity, $f_F$ the Fresnel knee frequency, $z$ the distance to the scattering screen, and $\lambda$ the wavelength of the observing frequency;

\begin{equation}
    V=f_F\sqrt{z\lambda\pi}.
	\label{eq:3}
\end{equation}

In the case of the model-fitting theoretical model used for SSSF analysis mode is given for the weak scintillation region, the formula can be represented by a theoretical temporal power spectrum $P(f)$ which is shown in equation~(\ref{eq:4}), where $f$ is the temporal frequency, $C=(2\pi r_e\lambda)^2$ is a constant related to observing wavelength $\lambda$, and $r_e$  is the electron radius.

\begin{equation}
    P(f)=C\int ^z _{-z}\frac{1}{V_p(z)}dz\int^{+\infty}_{-\infty}\Phi_{ne}F_{diff}F_{source}dq_y.
	\label{eq:4}
\end{equation}

In equation~(\ref{eq:4}), the spectrum of the electron-density fluctuations $\Phi_{ne}\propto q^{-\alpha}$, where $q=\sqrt{q^2_x+q^2_y+q^2_z}$ is the three-dimensional wave number. $F_{diff}$ is the Fresnel propagation filter and $F_{source}$ is the squared modulus of the source visibility \citep{manoharan1990determination, 2015SoPh..290.2539M, 2019SpWea..17.1114C}.\\

In some previous studies, the fitting parameters like $AR$ and $\alpha$ are often set to be a fixed number for simplicity, and only to do the model-fitting \citep{2000Oberoi} by adjusting the solar wind velocity $V$. In our IPS studies, a numerical optimization applies a weighted trust-region reflective least square (TRRLS) algorithm to optimize the four parameters (including $V$, $AR$, $\alpha$, and source size of the radio source) of IPS model, which are based on the physical constraints. The fitting parameters are obtained via fitting the theoretical model to the actual power spectrum. We used the $95\%$ confidence intervals as the error bars of the parameters. Moreover, the multi-initialization strategy applied will guarantee the fitting convergence in the global minimization. In order to obtain a good performance in the fitted velocity, we used the weighted nonlinear least-square fitting. The objective function of this method is:

\begin{equation}
    \min\,\frac{1}{N_m}\sum^{N_m}_{i=1}W(i)\|\lg(P_m(V,AR,\alpha,\theta_0)-\lg(P_{obs})\|^2_2.
	\label{eq:5}
\end{equation}

where $N_m$ is the number of points which will be used to carry out model-fitting, $W$ is weight, $P_m(V,AR,\alpha,\theta_0)$ and $P_{obs}$ are the IPS model and observed power spectra, respectively.\\

Fig.~\ref{fig:6} shows some examples of the SSSF model-fitting results of 3C\,286 with different time lengths, and the central observing frequency is set to be at 285\,MHz. The solid and dashed lines show the observed and model-fitting spectrum respectively. The time length of the power spectrum of Fig.~\ref{fig:6} (a), (b) and (c) is 20\,s, and that of Fig.~\ref{fig:6} (d) is 300\,s. Table~\ref{tab:3} are the parameters with error bars obtained by a data set of 300\,s, each velocity is obtained with a 20\,s time length. The corresponding value for data with 300\,s are: $V=598.2\pm10.2\,km/s$, $AR=1.1\pm0.3$, $\alpha= 2.7\pm0.5$, source size$=0.03\pm0.03\,''$.  The weighted mean value with scatter for each column of Table~\ref{tab:3} are: $V=620.9\pm4.1\,km/s$, $AR=1.1\pm0.1$, $\alpha=2.8\pm0.2$ and source size$=0.04\pm0.01\,''$. According to ISEE, the solar wind velocity obtained from observations of 3C\,286 on 14 November 2017 was 609\,km/s. It can be concluded that resulting solar wind velocity with the data of 20\,s integration is consistent with the ones with the time length of 300\,s. That means the model-fitting method developed is applicable, therefore, with the high sensitivity of FAST, the solar wind velocity can be derived from a 20\,s observation for this source that day.\\

\begin{figure}
	\includegraphics[width=\columnwidth]{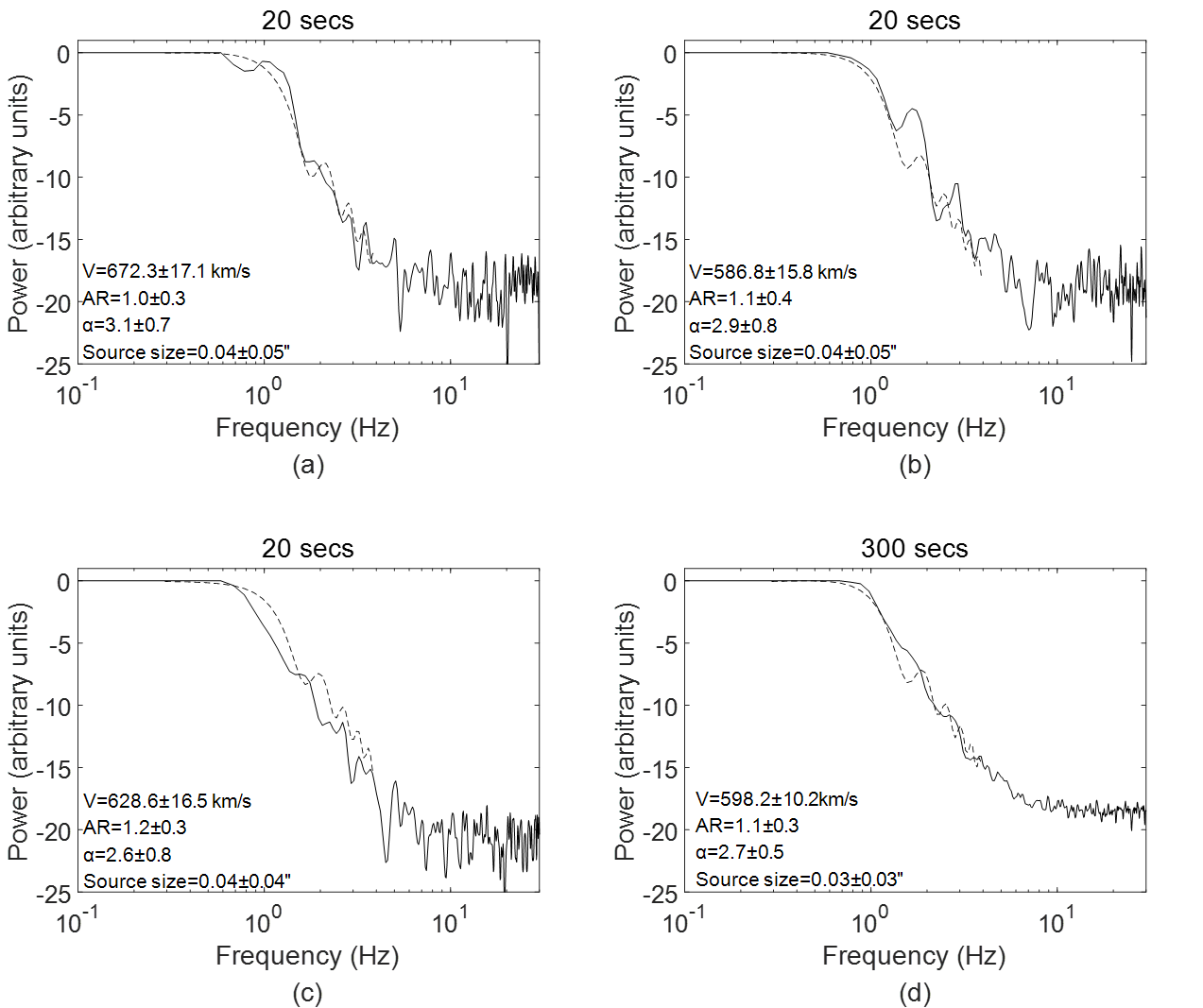}
    \caption{The model-fitting example of source 3C\,286 with SSSF analysis mode, the central observing frequency is 285\,MHz. The solid and dashed lines show the observed and fitting spectra respectively. (a), (b), (c) are the results for the time length of 20\,s, and (d) is that of 300\,s. }
    \label{fig:6}
\end{figure}

\begin{table}
	\centering
	\caption{Shown here are the fitted parameters with error bars for the 300\,s observation divided up into 20\,s intervals using the SSSF analysis method on the central frequency of 285\,MHz with a 10\,MHz bandwidth. We used the $95\%$ confidence intervals as the error bars of the parameters. The last row demonstrates the weighted mean values with scatter for each column.}
	\label{tab:3}
	\begin{tabular}{lccr} % four columns, alignment for each
		\hline
		$V$ & AR & $\alpha$ & Source size\\
        $(km/s)$& & &(arcsec)\\
		\hline
		626.2$\pm$14.5&	  1.1$\pm$0.4&	2.7$\pm$0.7&   0.04$\pm$0.04\\
        636.4$\pm$17.5&	  1.1$\pm$0.3&	2.8$\pm$0.8&   0.04$\pm$0.05\\
        552.4$\pm$11.4&	  1.1$\pm$0.2&	3.0$\pm$0.5&   0.01$\pm$0.09\\
        672.3$\pm$17.1&	  1.0$\pm$0.3&	3.1$\pm$0.7&   0.04$\pm$0.05\\
        623.3$\pm$20.6&   1.1$\pm$0.4&	2.9$\pm$0.7&	0.01$\pm$0.12\\
        663.4$\pm$18.2&   1.0$\pm$0.3&	3.3$\pm$0.8&	0.01$\pm$0.19\\
        586.8$\pm$15.8&   1.1$\pm$0.4&	2.9$\pm$0.8&	0.04$\pm$0.05\\
        749.1$\pm$17.1&   1.0$\pm$0.3&	3.1$\pm$0.7&	0.04$\pm$0.05\\
        577.2$\pm$20.6&   1.0$\pm$1.0&	1.6$\pm$0.9&	0.04$\pm$0.05\\
        628.6$\pm$16.5&   1.2$\pm$0.3&	2.6$\pm$0.7&	0.04$\pm$0.04\\
        624.7$\pm$14.7&   1.0$\pm$0.4&	2.7$\pm$0.6&	0.04$\pm$0.04\\
        647.6$\pm$13.1&   1.0$\pm$0.3&	3.3$\pm$0.5&	0.01$\pm$0.14\\
        579.5$\pm$11.9&   1.0$\pm$0.4&	2.5$\pm$0.5&	0.04$\pm$0.03\\
        560.7$\pm$14.6&   1.1$\pm$0.5&	2.4$\pm$0.7&	0.04$\pm$0.04\\
        642.3$\pm$22.9&   0.9$\pm$0.7&	1.8$\pm$1.2&	0.07$\pm$0.05\\
        \hline
        620.9$\pm$4.1&    1.1$\pm$0.1&  2.8$\pm$0.2&    0.04$\pm$0.01\\
		\hline
	\end{tabular}
\end{table}

The low-frequency band of FAST UWB receiver covers a wide frequency range from 270 to 800\,MHz, which is divided into 53 sub-bands, and each sub-band covers a frequency of 10\,MHz. Fig.~\ref{fig:7} shows the results of the four fitting parameters (V, AR, $\alpha$, and source size) obtained from each sub-band, the bands those affected by strong RFI are rejected. The time length for each result is 20\,s. The stars and solid lines are the fitting values and the error bars respectively.\\

\begin{figure}
	\includegraphics[width=\columnwidth]{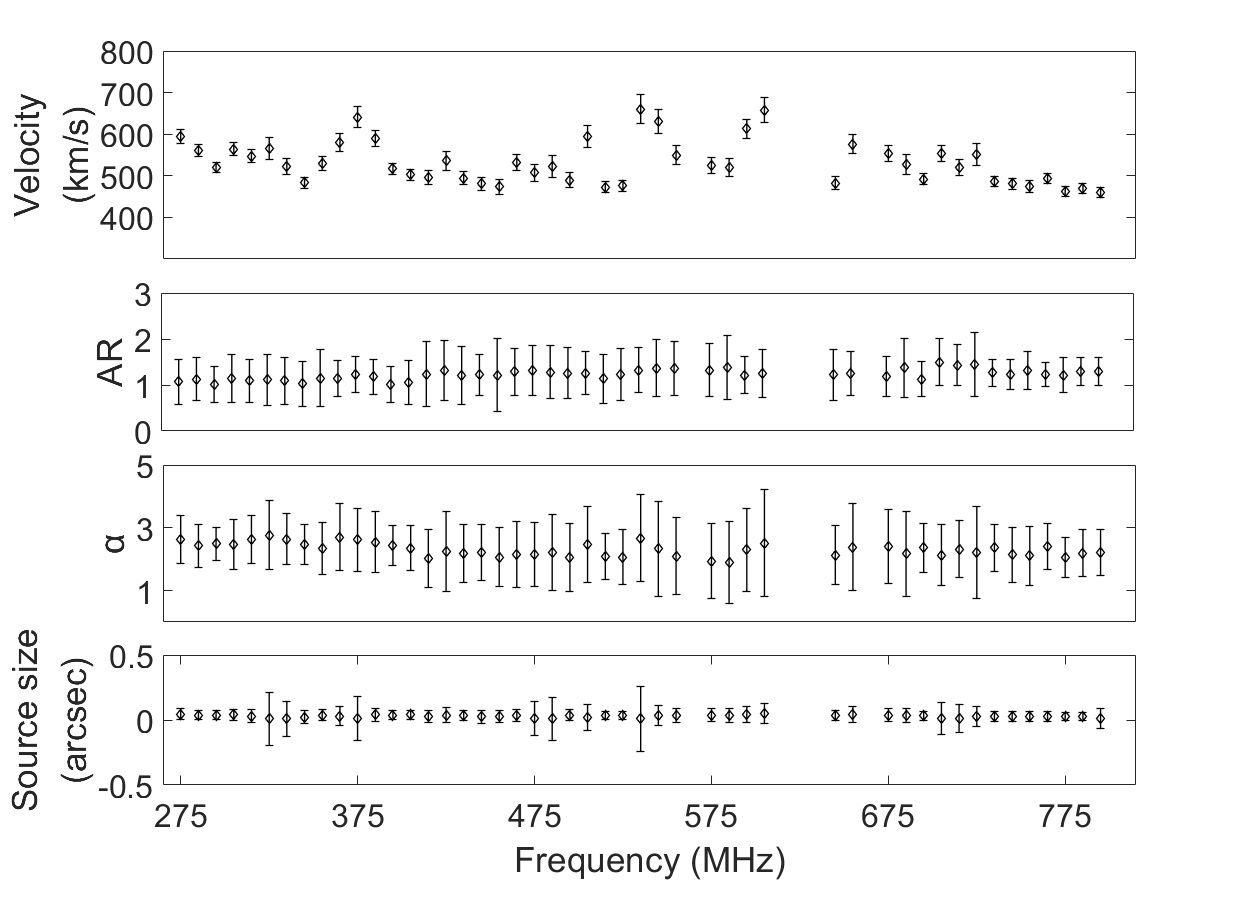}
    \caption{The solar wind parameters from the 10\,MHz bandwidth sub-bands using 20\,s time length. From top to down, the four parameters, velocity, AR, $\alpha$, and source size. The sub-bands affected by strong RFI are rejected. The stars and solid lines represent the fitting values and the error bars.
    }
    \label{fig:7}
\end{figure}

 The mean velocity in Fig.~\ref{fig:7} is 531.9\,km/s with a standard deviation of 52.7\,km/s, revealing velocity estimates  from about 600\,km/s at lower frequency to $\sim 450\,km/s$ at higher frequency, with a fractional variability of $\sim 9.9\%$ over a frequency range of $\sim 500$\,MHz, mostly in linear trend. The model-fitting values of AR, alpha, and the source-size component tend towards constants, which can also confirm that why some former studies set these three parameters in a fixed number \citep[e.g.][]{2000Oberoi}.  That means the parameters obtained from the model-fitting methods are reliable, and the new telescope FAST has very good advantages for investigating the inner heliosphere with IPS using both analyses types.\\

For SSDF analysis mode, the solar wind velocity is deduced from the first zero-crossing frequency of the N.C.S.  In equation~(\ref{eq:6}) $\lambda_1$ is the wavelength of the lower observing frequency, $z$ is the effective screen distance, $f_{zero}$ is the first zero-crossing frequency of N.C.S, $A$ is a correcting factor which varies slightly with the solar wind parameters and is always set to a constant, 1 \citep{1983A&A...123..191S, 1994JGG....46..835T, 2007ChJAA...7..712Z}.

\begin{equation}
    V=Af_{zero}\sqrt{z\lambda_1}.
	\label{eq:6}
\end{equation}

\begin{table}
	\centering
	\caption{SSSF and SSDF analyses modes velocity results with a duration time of 300\,s. The 1-3 columns are the frequency, velocity, and error, respectively.}
	\label{tab:4}
	\begin{tabular}{lccr} % four columns, alignment for each
		\hline
		Frequency & Velocity & Error \\
        (MHz)&(\,km/s) &(km/s)\\
		\hline
		305 &  593.4 &  10.2\\
        715 & 465.0 &  7.3\\
        305/715& 747.8 &  -\\

		\hline
	\end{tabular}
\end{table}

Fig.~\ref{fig:8} shows the result of the SSDF spectrum of 3C\,286 observed on 13 November 2017. The two observing frequencies are 305\,MHz and 715\,MHz. The time length of the dataset is 20\,s and the arrow shows the first crossing frequency $f_{zero}$  of the N.C.S. which is 2.05\,Hz, and the deduced solar wind velocity is 713.0\,km/s.  The velocities at the same time length from the SSSF analysis mode are 564.4$\pm$16.5\,km/s (305\,MHz) and 554.2$\pm$19.6\,km/s (715\,MHz) taken from Fig.~\ref{fig:7}, which is consistent with the velocity variation trend observed in Fig.~\ref{fig:7}. Table~\ref{tab:4} presents the velocity results of SSSF and SSDF analyses modes with a duration time of 300\,s. From column 1-3 shows the frequency, velocity, and error, respectively. It can be seen that the velocities deduced from 20\,s are consistent with the results of 300\,s. The velocity deduced from SSDF analysis mode is likely to be affected by the random velocity \citep{1994JGG....46..835T}, that may be the reason why the velocity of SSDF is higher than the velocity obtained from SSSF analysis mode. Or may be that the IPS is weak at that time, so the reliability of the velocity determination is reduced. But the cause of the difference of velocity still needs further study.
\\

\begin{figure}
	\includegraphics[width=\columnwidth]{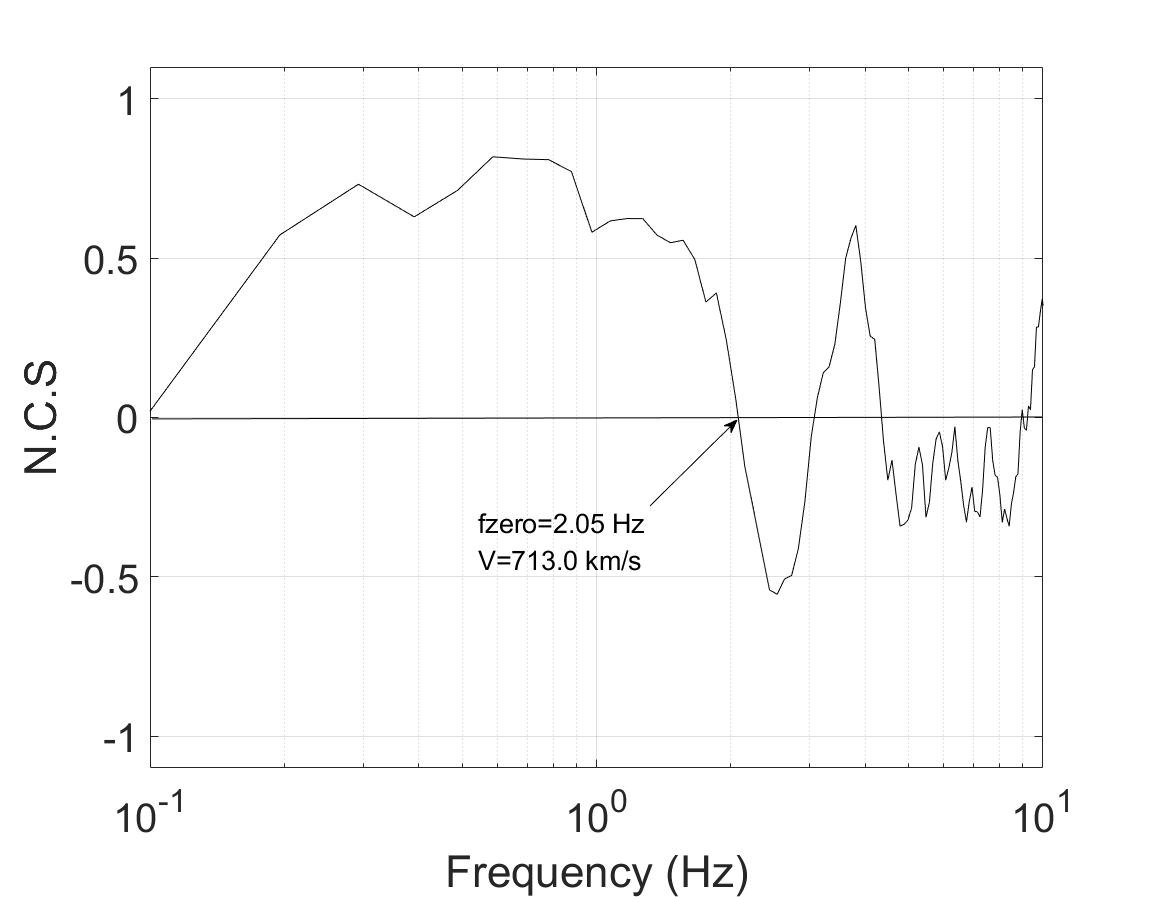}
    \caption{The power spectra of source 3C\,286 in SSDF analysis modes, which is observed on 13 November 2017. The two frequencies adopted here are 305\,MHz and 715\,MHz. The time length of this data set is 20\,s, and the arrow shows the $f_{zero}$ }
    \label{fig:8}
\end{figure}

\section{Discussions and Concluding Remarks}

There are four remarkable advantages to carry out IPS observations with the newly established telescope FAST. Firstly, FAST is incredibly sensitive to search for weak signals owing to its unique huge collecting area, which allows us to obtain the solar wind velocity information in a short observation of $\sim20\,s$. Since the IPS is a fast-changing phenomenon with a timescale of $\sim 1\,s$  \citep{1969ARA&A...7..619C}, the long integration is likely to obscure the scintillation phenomena. Secondly, the bandwidth can be chosen flexibly. Otherwise, if the adopted bandwidth is larger than the coherence scales, the scintillation will be smeared because the IPS is a frequency-dependent phenomenon \citep{little1968effect}. Thirdly, with the wide frequency coverage of the receiver and multi-channels, IPS observations in SSSF and SSDF analysis modes can be conducted with FAST simultaneously. The velocities derived from the two analysis modes provide a good supplement. In addition, IPS studies do not require large extra and dedicated observing time, for if the position of the source is appropriate we can take most calibrators as IPS targets which accommodate a symbiotic project.\\

Because of the wide frequency coverage of the UWB receiver, we can also study the effect of different frequency difference of SSDF analysis mode. Table~\ref{tab:5} shows the result of the comparison of different frequency differences $\Delta f$ ($\Delta f=f_2-f_1$). $f_1$  in Table~\ref{tab:5} is set to be 285 MHz. The five columns are two observing frequencies, $\Delta f$, the first zero-crossing frequency and the deduced solar wind velocity. It can be concluded from Table~\ref{tab:5} that, when $\Delta f$ is bigger than $30\%$ of  $f_1$,  then the deduced $f_{zero}$  is tend to be stable. The reason of this frequency differences still request some follow up study.\\

\begin{table}
	\centering
	\caption{The effect of different $\Delta f$. Column 1-5 shows the $f_1$, $f_2$, $\Delta f$, $f_{zero}$ and $V$ respectively}
	\label{tab:5}
	\begin{tabular}{lccccr} % four columns, alignment for each
		\hline
		%$f_1$&$f_2$&$\Delta f=f_2-f_1$&$f_{zero}$&$V$ \\
        $f_1$ & $f_2$ & $\Delta f=f_2-f_1$ & $f_{zero}$ & $V$\\
        $(MHz)$ & $(MHz)$ & $(MHz)$ & (Hz)& (km/s)\\
		\hline
		285 & 335 & 50&  3.516&  1265.1\\
        285 & 385 & 100& 2.148&  772.9\\
        285 & 435 & 150& 2.051&  737.9\\
        285 & 485 & 200& 2.051&  737.9\\
        285 & 535 & 250& 2.051&  737.9\\
        285 & 585 & 300& 2.051&	 737.9\\
        285 & 635 & 350& 2.051&	737.9\\
        285 & 685 & 400& 2.148&	772.9\\
        285 & 735 & 450& 2.051&	737.9\\
        285 & 785 & 500& 2.051&	737.9\\

		\hline
	\end{tabular}
\end{table}

Our preliminary results demonstrated that FAST has an outstanding potential to perform observations of IPS. By comparing the solar wind velocities calculated from the SSDF and SSSF analysis modes, the reliability of observation is proven. In general, taking advantage of the high sensitivity and wide frequency coverage as well as the multi-channel design of the UWB receiver, the observation time with FAST can significantly decrease on each IPS source, and the choice of frequencies is flexible to get rid of the RFI contaminated data.\\

We conclude that a larger number of target sources can be observed with FAST on a short time scale. therefore, the IPS data from FAST will widen the range of Solar-terrestrial space and improve the accuracy of estimation of solar wind velocity. This can be further utilized in space weather forecasts.\\

\section*{Acknowledgements}

This work is supported by the National Key $R \& D$ Program of China under grant number 2018YFA0404703, and the Open Project Program of the CAS Key Laboratory of FAST, NAOC, Chinese Academy of Sciences. Basic research program and project of Yunnan province of China (2019FB009). The authors thank all the staff of JLRAT, NAOC, for their help during the observations. We are grateful to Yu-Hai Qiu, Xi-Zhen Zhang, Cheng-Min Zhang, Ming Xiong (National Space Science Center, CAS), R. A. Fallow (ASTRON), M. Tokumaru (Solar-Terrestrial Environment Laboratory, Nagoya University, Nagoya, Japan) and Julio Mejia-Ambriz (SCiESMEX, Instituto de Geofisica, Unidad Michoacan, Universidad Nacional Autonoma de Mexico), for their helpful discussions. The authors also thank Yue Ma and Sivasankaran Srikanth (National Radio Astronomy Observatory, NRAO), for their help to polish the English. O. C. acknowledges the European Union Horizon 2020 research and innovation program under the Marie Sk{\l}odowska-Curie grant agreement 665593 awarded to the Science and Technology Facilities Council. M. M. B. acknowledges STFC in-house research funding to RAL Space under the STFC Astronomy Grants Panel Consolidated Grants Programme. This work made use of the data from FAST (Five-hundred-meter Aperture Spherical radio Telescope). FAST is a Chinese national mega-science facility, operated by the National Astronomical Observatories, Chinese Academy of Sciences.

%%%%%%%%%%%%%%%%%%%%%%%%%%%%%%%%%%%%%%%%%%%%%%%%%%
\section*{Data Availability}

The raw data underlying this article cannot be shared publicly due to the data policy of FAST. The intermediate process data will be shared on reasonable request to the corresponding author.



\bibliographystyle{mnras}
\bibliography{ips} % if your bibtex file is called example.bib

%\begin{thebibliography}{99}
%\end{thebibliography}

%%%%%%%%%%%%%%%%%%%%%%%%%%%%%%%%%%%%%%%%%%%%%%%%%%

%%%%%%%%%%%%%%%%% APPENDICES %%%%%%%%%%%%%%%%%%%%%

%\appendix

%%%%%%%%%%%%%%%%%%%%%%%%%%%%%%%%%%%%%%%%%%%%%%%%%%

% Don't change these lines
\bsp	% typesetting comment
\label{lastpage}
\end{document}